\documentclass[journal]{IEEEtranTIE}

\IEEEoverridecommandlockouts                              

\usepackage{graphicx}
\usepackage{cite}
\usepackage{picinpar}
\usepackage{amsmath}
\usepackage{url}
\usepackage{flushend}
\usepackage[latin1]{inputenc}
\usepackage{colortbl}
\usepackage{soul}
\usepackage{multirow}
\usepackage{pifont}
\usepackage{color}
\usepackage{alltt}
\usepackage[hidelinks]{hyperref}
\usepackage{enumerate}
\usepackage{siunitx}
\usepackage{breakurl}
\usepackage{epstopdf}
\usepackage{pbox}
\usepackage{amssymb,amsfonts,mathtools, comment}
\usepackage{array}
\newcolumntype{M}[1]{>{\centering\arraybackslash}p{#1}}
\graphicspath{{./images/}}
\usepackage{wrapfig}
\usepackage{subcaption}
\usepackage{soul}
\usepackage{tikz}

\newcommand\norm[1]{\left\lVert#1\right\rVert}
\newcommand{\colvec}[2][.8]{%
  \scalebox{#1}{%
    \renewcommand{\arraystretch}{.8}%
    $\begin{smallmatrix}#2\end{smallmatrix}$%
  }
}

\title{\LARGE \bf
Robust Optimal Safe and Stability Guaranteeing Reinforcement Learning Control for Quadcopter
}

\author{
	\vskip 1em

	Sanghyoup Gu, \emph{Student Member, IEEE};
	Ratnesh Kumar, \emph{Fellow, IEEE}

	\thanks{
	
		Sanghyoup Gu and Ratnesh Kumar are with the Electrical and Computer Engineering Department, Iowa State University, Ames, IA 50014, USA. (e-mails: rndk9004,rkumar@iastate.edu). 
	}
}

\begin{document}
\newcommand\copyrighttext{%
 \footnotesize This work has been submitted to the IEEE for possible publication. Copyright may be transferred without notice, after which this version may no longer be accessible.}
\newcommand\copyrightnotice{%
\begin{tikzpicture}[remember picture,overlay]
\node[anchor=south,yshift=10pt] at (current page.south) {\fbox{\parbox{\dimexpr\textwidth-\fboxsep-\fboxrule\relax}{\copyrighttext}}};
\end{tikzpicture}%
}

\maketitle
\copyrightnotice
\thispagestyle{empty}
\pagestyle{empty}

\begin{abstract}
Recent advances in deep learning have provided new data-driven ways of controller design to replace the traditional manual synthesis and certification approaches. However, employing neural networks (NN) as controllers presents its own challenge: that of certifying safety, stability and robustness, due to their inherent complex nonlinearity, and while NN controllers have demonstrated high performance in complex systems, they often lack formal guarantees. This issue is further accentuated for critical applications such as unmanned aerial vehicles (UAVs), whereas a lack of safety or stability or robustness assurance raises the risk of critical damage or even complete failure. In this study, we improve a Robust, Optimal, Safe, and Stability Guaranteed Control (ROSS-GC) method of \cite{Soumya} to design an NN controller for a quadcopter flight control. The approach ensures robust closed-loop system stability by finding a robust Lyapunov function and providing a safe initial state domain that remains invariant under the control and also guarantees asymptotic stability to an equilibrium within it even under random parametric variability. Robust stability guaranteeing constraints are derived from the sector bound of the system nonlinearity and of its parametric variations, in the form of a Lipschitz bound for NN control. The control performance is next optimized by searching over the class of controllers having the said Lipschitz bound to minimize the reference tracking error together with the control costs. 
\end{abstract}


\section{INTRODUCTION}

\IEEEPARstart{I}{n} recent decades, neural networks (NNs) based data-driven approach has achieved great progress, finding applications across diverse fields, including tne promising use case of the design of NN controllers for dynamical systems, where a neural network maps state observations to actions. Unlike conventional feedback controllers, which rely either on predefined structures with domain-specific expertise for their tuning, or the control Lyapunov functions that require their exploration, NN controllers are trained through data-driven approaches. For instance, supervised imitation learning uses images from a human operator and trains a neural network to imitate those actions \cite{IM1_Fan_2020},\cite{ImitationLearning2}. Similarly, in reinforcement learning (RL), policy networks are optimized to maximize the expected cumulative rewards from control sequences \cite{RLBook}.

While NN controllers have demonstrated good performance in many applications, their deployment in safety-critical systems, such as unmanned aerial vehicles (UAVs), remains limited due to the lack of formal guarantees of safety, stability, and robustness. Safety corresponds to operting in the safe region of the state-space, stability requires convergence to reference, while robustness if the ability to withstand disturbances, sensor noise, or model uncertainties without loss of safety or stability. In real-world environments of uncertainties, robust control is essential for reliable operation, which again requires a formal certification that, in general, is lacking for NN controllers.

This paper enhances and applies the Robust, Optimal, Safe, and Stability-Guaranteed Control (ROSS-GC) method of \cite{Soumya} to design and train an NN controller for a quadcopter, a widely used form of UAV. The ROSS-GC method consists of two stages: First, it extracts the Lyapunov-based constraints to ensure robust safety and stability for a closed-loop nonlinear system with parametric variability, as well as a safe initial state domain that is invariant, and guarantees converge to equilibrium within in spite of the parametric variations, utilizing the sector bounds of system nonlinearities and its parametric variations, and imposing Lipschitz bound for the neural network (so their output magnitude is bounded by the input magnitude and the Lipschitz bound gain). The maximal permissible robust, safe, and stabilizing Lipschitz bound for the NN controller and the maximal initial safe state domain ensuring asymptotic robust stability under bounded parametric uncertainty are derived by iteratively analyzing a linear matrix inequality (LMI) certificate. Second, it finds an optimal NN controller satisfying the Lipschitz bound obtained in the first step to minimize control cost and reference trajectory deviation error using reinforcement learning.

The original ROSS-GC method \cite{Soumya} employs the infinity norm to derive the Lipschitz bound for the NN controller. However, this approach often yields overly conservative estimates for the NN Lipschitz constant, particularly for NNs with many neurons, making it impractical. In this paper, we propose an enhancement, replacing the infinity norm with the Euclidean norm, demonstrating that it still satisfies the same LMI constraints for the existence of a robust Lyapunov function, in spite of reducing the above-mentioned conservativeness. Furthermore, we utilize a semidefinite programming-based method for the estimation of the Lipschitz constant of a neural network \cite{LipSDP} that further reduces the conservativeness of the method in \cite{Soumya} (the result is a $>\!\!2400$-fold reduction in conservativeness for the quadcopter example considered), enabling the practicality of robust, safety, and stability guaranteeing NN controller design.

\subsection{Related works on learning-based control}
In the imitation learning setting of \cite{IM1_Fan_2020}, a human expert remotely controls a drone equipped with a camera to follow a moving object on the ground. The video recording of the human expert collected during this process is subsequently used to train a neural network to imitate the expert's command for the corresponding raw inputs. In the reinforcement learning (RL) setting of \cite{Hwangbo_2017}, a deterministic on-policy gradient method was applied to train an NN controller for a quadcopter to recover its position under harsh disturbances, demonstrating its stabilizing ability, but without any formal guarantee. Studies in \cite{Koch_2019}, comparing deep RL methods such as Deep Deterministic Policy Gradient (DDPG), Trust Region Policy Optimization (TRPO), and Proximal Policy Optimization (PPO) with traditional PID controllers have found that RL-trained NN controllers outperform their PID counterparts in attitude control tasks. These findings highlight the potential of NN controllers for quadcopter control.

However, a critical limitation of these approaches is their lack of formal guarantee of safety, stability, and robustness to environmental and model uncertainties. To address this, researchers have explored methods to robustify the training of NN controllers. Domain randomization, for example, has been employed to account for uncertainties by randomizing model/disturbance parameters during the training. This method enabled a quadcopter to land autonomously at a target point, with results showing that the adequately randomized domains allowed the trained policy to generalize to new environments without any additional training \cite{Polvara_2020}. Similarly, the Robust Markov Decision Process (RMDP) was used in \cite{Aditya_2021} to address environmental and model uncertainties, where RMDP maximizes cumulative rewards under worst-case scenarios. Quadcopter attitude control trained using RMDP demonstrated satisfactory trajectory tracking in uncertain environments \cite{Aditya_2021}. While these methods improve robustness, they still lack any formal guarantee. 

From a control theory perspective, some studies have analyzed NN controllers to ensure safety, stability, and robustness. A common approach involves employing a Lyapunov function, which provides a sufficient condition for asymptotic stability. One approach is to model the Lyapunov function as well as the controller as neural networks and then train them to certify closed-loop stability \cite{LyapunovStableNNControl, NeuralLyapunovControl, NeuralLyapunovControlUnknown, NeuralLyapunovControlDiscrete, NeuralLyapunovControlOutput}. The setup consists of a \textit{learner} and a \textit{verifier}, where the \textit{verifier} validates the Lyapunov constraint and the \textit{learner} updates the weights of the Lyapunov and the controller NNs. The \textit{verifier}, using techniques such as mixed-integer programming (MIP) or satisfiability modulo theories (SMT), identifies counterexamples violating stability constraints. If no counterexamples are found, the Lyapunov constraint is considered satisfied. The \textit{learner} uses counterexamples to tune the weights in the Lyapunov function and the controller. Simulation in \cite{LyapunovStableNNControl} showed that the trained controller successfully stabilized a quadcopter in a hovering position. However, such a validation of the Lyapunov constraint is undecidable in general, and the empirical approaches are sound but not complete, and also, the computational complexity of MIP/SMT solvers during the training process poses challenges for adoption to large-scale problems.    

Another control-theoretic approach leverages quadratic constraints (QCs) to bound the output of NN controllers, abstracting activation function properties \cite{Jin_2018, SafetyVerification, StabilityAnalysisUsingQC, Soumya}. The method in \cite{SafetyVerification} formulated an LMI constraint to approximate the input-output map of a neural network. When the input domain with disturbance is given, safety and robustness are proved by showing that the output approximation lies within a safe region. To prove robust closed-loop stability with NN controller, \cite{StabilityAnalysisUsingQC} derived LMI constraint from integral quadratic constraints (IQCs) that capture the input-output behavior of sector-bounded nonlinear components of the system and QCs that bound the nonlinear activation function in NN controller. However, this method is mainly suited for static networks, and also is difficult to integrate into the training process. An approach in \cite{Jin_2018} derives QCs from sector bounds of smooth functions to guarantee input-output stability. Extending this idea, \cite{Soumya} developed QCs for Lipschitz-bounded NN controllers, sector-bounded nonlinear components of the system, and its variability, deriving LMI constraints for robust Lyapunov stability. From the proposed LMI, a Lipschitz constraint for an NN controller that can maintain robust stability is obtained. Then, the NN controller is trained while maintaining that its Lipschitz bound lies below the derived constraint. 

The methods using QCs represent nonlinear activation functions using constraints that capture an activation function's property, such as the resulting Lipschitz bound. To obtain a practical solution, however, it is important to obtain non-conservative Lipschitz bound constraints, and we propose certain such enhancements in the paper, along with their application to quadcopter control for the first time. 
Our work leverages these advancements of robust, safe, and stability-guaranteed NN control design for the navigation of quadcopters for optimal reference trajectory tracking. Our contribution further includes enhancing the method of \cite{Soumya} to allow for a less conservative Lipschitz bound by using 2-norm in place of infinity-norm while preserving the key results of robust stability and by estimating the Lipschitz bound of a NN controller using a less conservative method involving semi-definite programming.

\subsection{Organization}
The remainder of this paper is organized as follows. Section \uppercase\expandafter{\romannumeral2} briefly discusses quadcopter dynamics. Section \uppercase\expandafter{\romannumeral3} reviews the robust optimal safe and stability-guaranteed control (ROSS-GC) method of \cite{Soumya}. Section \uppercase\expandafter{\romannumeral4} presents the enhancement and application of ROSS-GC for a quadcopter. Section \uppercase\expandafter{\romannumeral5} provides the simulation results of the trained NN controller. Section \uppercase\expandafter{\romannumeral6} then provides a conclusion.

\subsection{Notation}
For a vector $x \in \mathbb{R}^n$, $x_i$ denotes its $i$-th element. $\norm{x}_p$ denotes vector's $p$ norm when $p \geq 1$ and $\norm{x}_\infty$ denotes the infinity norm, $diag(x)$ denotes a diagonal matrix whose $i$-th diagonal element is $x_i$. For a matrix $M \in \mathbb{R}^{m \times n}$, $M_{ij}$ denotes its element in $i$-th row and $j$-th column. Operators $\leq, <, \geq$, and $>$ on matrices or vectors are elementwise operations. For simplicity of notation, a symmetric matrix is denoted by $P = \left[\begin{array}{cc} P_{11}&*\\P_{21}&P_{22} \end{array}\right]$. $\mathbb{S}, \mathbb{S}^{\geq 0}$, and $\mathbb{S}^{+}$ represent the set of symmetric, positive semi-definite, and positive definite matrices, respectively.

\section{Quadcopter Model}
In this section, quadcopter dynamics is briefly discussed. Quadcopter dynamics consists of translation and rotation movement. In the case of translation movement, gravitation, thrust, and drag force, which is proportional to the velocity, impact the translational acceleration. 

\begin{align}  
m\left[\begin{array}{c}
\ddot{x} \\ \ddot{y} \\ \ddot{z} \end{array}\right] = \left[\begin{array}{c} 0\\0\\-g \end{array}\right]+\boldsymbol{R}\left[\begin{array}{c} 0\\0\\T \end{array}\right] - \boldsymbol{D} \left[\begin{array}{c} \dot{x} \\ \dot{y} \\ \dot{z} \end{array}\right], \label{eq1}
\end{align}
where $[x, y, z]^T \in \mathbb{R}^3$ is the position in the inertial frame, $m$ is the mass, $g$ is the gravitational acceleration, $T$ is the thrust produced from the propellers, $\boldsymbol{R}\in SO(3)$ is the rotation matrix from the body frame to the inertial frame [$SO(3)$ is special orthogonal group, i.e., $SO(3) = \{\boldsymbol{R} \in \mathbb{R}^3 | \boldsymbol{R}^T\boldsymbol{R} = I, det(\boldsymbol{R})=1\}$], and $\boldsymbol{D} = diag([D_x, D_y, D_z]^T)$ is the diagonal drag coefficient matrix. The rotation matrix $\boldsymbol{R}$ is given by the equality (\ref{eq2}) involving the Euler angles $[\phi,\theta, \psi] \in \mathbb{R}^3$  of roll, pitch, and yaw between the inertial and the body frames:
\begin{align}  
 \boldsymbol{R} = \left[\begin{array}{ccc} C_{\psi}C_{\theta}&C_{\psi}S_{\theta}S_{\phi}-S_{\psi}C_{\phi}&C_{\psi}S_{\theta}C_{\phi}+S_{\psi}S_{\phi}\\S_{\psi}C_{\theta}&S_{\psi}S_{\theta}S_{\phi}+C_{\psi}C_{\phi}&S_{\psi}S_{\theta}C_{\phi}-C_{\psi}S_{\phi}\\-S_{\theta}&C_{\theta}S_{\phi}&C_{\theta}C_{\phi} \end{array}\right], \label{eq2}
\end{align}
in which $S_a$ and $C_a$ indicates $sin(a)$ and $cos(a)$, respectively. \\

In the case of rotation movement, the angular acceleration in the body frame is produced by the centripetal force and the torque. The gyroscopic force is neglected because it is much smaller than the other two components.  

\begin{align}  
\left[\begin{array}{c}
\dot{p} \\ \dot{q} \\ \dot{r} \end{array}\right] = \left[\begin{array}{c} (I_{yy}-I_{zz})qr/I_{xx}\\(I_{zz}-I_{xx})pr/I_{yy}\\(I_{xx}-I_{yy})pq/I_{zz}\end{array}\right]+\left[\begin{array}{c} \tau_{\phi}/I_{xx} \\ \tau_{\theta}/I_{yy} \\ \tau_{\psi}/I_{zz} \end{array}\right] , \label{eq3}
\end{align}
where $[p,q,r]^T \in \mathbb{R}^3$ is the angular velocity in the body frame, $[I_{xx},I_{yy},I_{zz}]$ are the three moment of inertia components, and $[\tau_{\phi}, \tau_{\theta}, \tau_{\psi}]^T$ is the torque produced from the propellers. \\  

The angular velocity in body frame and Euler angle rate are related by the equality (\ref{eq4}): 

\begin{align}  
\left[\begin{array}{c} \dot{\phi} \\ \dot{\theta} \\ \dot{\psi} \end{array}\right] =  \left[\begin{array}{ccc} 1&S_{\phi}T_{\theta}&C_{\phi}T_{\theta}\\0&C_{\phi}&-S_{\theta}\\0&\frac{S_{\phi}}{C_{\theta}}&\frac{C_{\phi}}{C_{\theta}}\end{array}\right] \left[\begin{array}{c} p \\ q \\ r \end{array}\right]. \label{eq4} 
\end{align}

In this work, an X-shape quadcopter is considered, for which the thrust and torque values as generated by the individual propeller's rotations and the drag constants are given by,  

\begin{align}  
\left[\begin{array}{c} T\\\tau_{\phi}\\\tau_{\theta}\\\tau_{\psi}\end{array}\right] = 
\left[\begin{array}{cccc} k_t&k_t&k_t&k_t\\-\frac{lk_t}{\sqrt{2}}&-\frac{lk_t}{\sqrt{2}}&\frac{lk_t}{\sqrt{2}}&\frac{lk_t}{\sqrt{2}}\\-\frac{lk_t}{\sqrt{2}}&\frac{lk_t}{\sqrt{2}}&\frac{lk_t}{\sqrt{2}}&-\frac{lk_t}{\sqrt{2}}\\k_d&-k_d&k_d&-k_d\end{array}\right] \left[\begin{array}{c} \omega_1^2\\\omega_2^2\\\omega_3^2\\\omega_4^2\end{array}\right], \label{eq5}
\end{align}
where $\omega_i$ is the angular speed of the  $i$th rotor, $l$ is the length of each arm of the quadcopter, $k_t$ is the thrust constant, and $k_d$ is the drag constant. A more detailed explanation for quadcopter dynamics modeling can be found in \cite{Modeling}.

\section{Robust Optimal Safe and Stability Guaranteed Control Method}
We review the Robust Optimal Safe and Stability Guaranteed Control (ROSS-GC) method proposed in \cite{Soumya} to certify the robust safety and asymptotic stability of neural network (NN)-controlled closed-loop systems with random parameter variations. 

\subsection{System and its control framework}
The system under consideration consists of nonlinear dynamics with parameter variations, representing model uncertainties or disturbances:
\begin{align}
 &\dot{s}(t) = f(s(t),u(t),\alpha(t)), \label{eq6} \\ 
 &u(t) = \pi(s(t)) = \pi_0(s(t)) + \pi_{NN}(s(t)) \nonumber 
\end{align}
\noindent where $s(t)\in \mathbb{R}^n, u(t)\in \mathbb{R}^m$, and $\alpha(t)\in \mathbb{R}^d$ are state, control input, and parameter variation, respectively; $f: \mathbb{R}^n \times \mathbb{R}^m \times \mathbb{R}^d \rightarrow \mathbb{R}^n$ denotes the nonlinear plant dynamics assumed to be locally continuously differentiable; $\pi: \mathbb{R}^n \rightarrow \mathbb{R}^m$ denotes a state-feedback control policy that is viewed as a composition of a nominal $\pi_0(\cdot)$ and an NN $\pi_{NN}(\cdot)$ control.

An equilibrium point is defined as the state where the system dynamics remain unchanged ($\dot{s}=0$). In this study, the equilibrium point is assumed invariant under parameter variations, and without loss of generality, the coordinates are shifted such that the equilibrium point aligns with the origin ($s^* = \boldsymbol{0}$). The system (\ref{eq6}) is deemed robustly asymptotically stable if, starting from an initial state within a certain neighborhood of the equilibrium, it converges to the equilibrium point in spite of the parametric variations, assumed to be contained within a bounded domain. 

\begin{figure} [htbp]
\vspace*{-.2in}
\centerline{\includegraphics[width=0.4\textwidth]{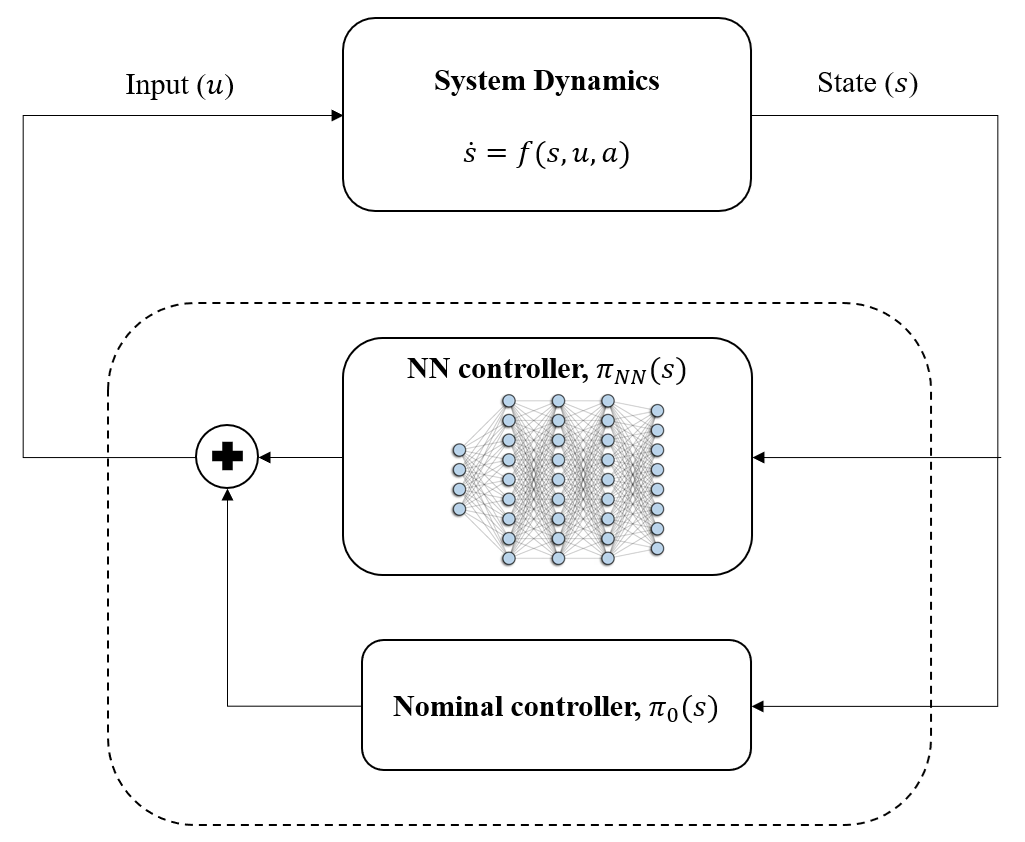}}
\caption{Controller with nominal vs. NN components}
\label{Control_architecture}
\end{figure} 
When parameter variations are zero, the system is referred to as the nominal system, and the corresponding control is termed the nominal control, and is designed to stabilize the nominal system linearized around the equilibrium point. The NN control is thus viewed as a perturbation around the nominal control to ensure the stability of the overall nonlinear parameter varying system (and not just the linearized nominal system). Fig \ref{Control_architecture} illustrates this structure of the control architecture, which integrates both the nominal and NN controllers. Both the nominal and NN controllers operate as state-feedback controllers, with the plant receiving control command as the sum of their respective outputs.

\noindent \textbf{Definition 1 (Lyapunov function \cite{Nonlinear})} 
For a dynamical system with an equilibrium point $s^*$, a continuously differentiable function $\mathcal{V}: \mathcal{S} \rightarrow \mathbb{R}_{\geq0}$, where $\mathcal{S} \subset \mathbb{R}^n$ is a compact domain such that $s^* \in$ \textbf{int}$(\mathcal{S})$, is a Lyapunov function and $\mathcal{S}$ is region-of-stability (RoS) if:
\begin{align}
\mathcal{V}(s)>0, \dot{\mathcal{V}}(s)<0, \forall s \in \mathcal{S}\setminus s^*; \nonumber \\
\mathcal{V}(s^*) = \dot{\mathcal{V}}(s^*) = 0, \nonumber  
\end{align}
where \textbf{int}$(\mathcal{S})$ denotes the interior of $\mathcal{S}$.

In \cite{Soumya}, a feasibility condition for the existence of a Lyapunov function comes from two quadratic constraints (QC) obtained from: 1) NN controller Lipschitz bound, and 2) system's nonlinear and parameter variation (NPV) sector bound. As demonstrated in \cite{Jin_2018}, Lipschitz continuous functions, such as the NNs, exhibit QCs derived from their bounded partial derivatives within a specified domain. Similarly, sector bounded NPVs imply another set of QCs. By combining these QCs, \cite{Soumya} proposed a Linear Matrix Inequality (LMI) constraint to be able to certify the existence of a robust Lyapunov function, as discussed in the following.  
\\

\noindent The ROSS-GC method of \cite{Soumya} consists of two stages:\\

\noindent \textbf{1. Extraction of maximal Lipschitz bound, robust and safe region of stability, and nominal control}: From the given system dynamics with bounds for parameter variations and a safety domain of operation, an LMI constraint for the existence of robust Lyapunov function is developed, and using which, the maximal RS-RoS, the maximal Lipschitz bound for the NN controller, and a state-feedback gain of a nominal controller are derived, based on the iterative feasibility check of the LMI constraint. \\

\noindent \textbf{2. Optimal NN Controller Training}: A NN controller is then trained using reinforcement learning while ensuring that its Lipschitz constant remains within the bound established in the first stage, and the cost of control, as well as the deviation of the state trajectory from a desired reference, is minimized. \\

The approach in \cite{Soumya} used the infinity norm for defining Lipschitz continuity and deriving QCs associated with the NN control. Further, the Lipschitz constant of the NN controller was estimated using the maximum absolute row sum norm of its weight matrices. However, this approach often yields an overly conservative estimate for the Lipschitz bound, particularly for high-dimensional networks, making it impractical. To address this, to be able to make the approach of \cite{Soumya} practical, here we propose replacing the infinity norm with the Euclidean (L2) norm, proving that the feasibility of the LMI constraint remains preserved under this change. Additionally, we employ a semidefinite programming-based method\cite{LipSDP} to obtain a tighter Lipschitz bound estimate, resulting in a more practical, relaxed Lipschitz bound for the NN controllers.

\subsection{Lipschitz bounded control with New enhancements}
A neural network (NN) that we use for control is a Lipschitz operator, meaning its output magnitude is bounded by the product of the input magnitude and a certain constant that depends on the NN parameters. This Lipschitz continuity of an NN can be defined using any norm; in this work, we adopt the L2-norm for the Lipschitz continuity, replacing the $\infty$-norm used in \cite{Soumya}. This change preserves all the results of \cite{Soumya}, yet makes the Lipschitz bound much less conservative, and so far more easier to satisfy.\\

\noindent \textbf{Definition 2 (Lipschitz continuity)} 
A function $\pi$ is Lipschitz continuous over a domain $\mathcal{S}$ if there exists a constant $L>0$ such that:
\begin{align} 
\norm{\pi(x) - \pi(y)}_2 \leq L\norm{x-y}_2, \forall x,y \in \mathcal{S}. \label{eq7}
\end{align}
\noindent where $L$ is termed the Lipschitz bound constant. The class of all control maps $\pi(\cdot)$ that are $L$-Lipschitz (Lipschitz bounded with bound $L$) is denoted by $\Pi_L$\\

As shown in \cite[Proposition 1]{Soumya}, each output of an $L$-Lipschitz NN can be expressed as a linear combination of scalar functions weighted by the input elements, with partial derivatives of the scalar functions bounded between $\pm L$. As shown next, this essential result remains valid even when using the L2-norm based definition of Lipschitz continuity, preserving all the ensuing results and the framework of \cite{Soumya}. From the definition of Lipschitz continuity (\ref{eq7}), we have:  
\begin{align} 
&\norm{\pi_{NN}(s_1)-\pi_{NN}(s_2)}_2 \leq L\norm{s_1 - s_2}_2 \nonumber \\
                                     & = L\sqrt{\sum_{j=1}^{n}{\lvert s_{1,j}-s_{2,j} \rvert}^2} 
                                     \leq L\sum_{j=1}^{n}|s_{1,j}-s_{2,j}|, \forall s_1, s_2 \in \mathbb{R}^n \nonumber      
\end{align}
where the last inequality follows from the triangular inequality. Also, since the infinity-norm is always less than the 2-norm: 
\begin{align} 
\norm{\pi_{NN}(s_1)-\pi_{NN}(s_2)}_{\infty} \leq \norm{\pi_{NN}(s_1)-\pi_{NN}(s_2)}_2, \nonumber
\end{align} 
we can conclude the same result as in the proof step of \cite[Proposition 1]{Soumya} even with the use of L2-norm to define Lipschitz continuity (as opposed to the $\infty$-norm): 
\begin{align} 
\norm{\pi_{NN}(s_1)-\pi_{NN}(s_2)}_{\infty} &\leq L\sum_{j=1}^{n}|s_{1,j}-s_{2,j}|. \nonumber      
\end{align}
It then follows that there exists scalar functions: $\delta_{ij}: \mathcal{S} \times \mathcal{S} \rightarrow [-L,L], \forall i \in \{1,...,m\}, \forall \in \{1,...,n\}$ such that $\forall s_1,s_2 \in \mathcal{S}$: 
\begin{align} 
\pi_{NN}(s_1) - \pi_{NN}(s_2) = \left[ \begin{array}{c} \sum_{j=1}^{n}\delta_{1j}(s_1,s_2)\big(s_{1,j}-s_{2,j}\big) \\ : \\ \sum_{j=1}^{n}\delta_{mj}(s_1,s_2)\big(s_{1,j}-s_{2,j}\big) \end{array} \right], \nonumber
\end{align}
which again is the same core result in the proof of \cite[Proposition 1]{Soumya}. Therefore, the QC of \cite[Proposition 1]{Soumya} is still valid even under the use of L2-norm based Lipschitz continuity. Note the change in norm, results in a new version of the input domain implied by the state domain $\mathcal{S}$ that is used below to obtain a local sector bound for system nonlinearity and parameter variability: 
\begin{align} 
\mathcal{U}_{L,\mathcal{S}} := \{ u_{NN} \in \mathbb{R}^m | & \exists s \in \mathcal{S}, \pi_{NN}(s) = u_{NN}, \nonumber \\
                                                            & \norm{u_{NN}}_2 \leq L\norm{s}_2 \}. \label{eq8}
\end{align}

A second enhancement we introduce is the manner in which the Lipschitz bound of an NN is estimated. Neural networks are composed of linear operations of weighted sums and nonlinear activations. Under the L2-norm based definition of Lipschitz continuity utilized above, the Lipschitz constant for the linear operators is simply determined by the largest singular values of the corresponding weight matrices \cite{NNLipschitz}. On the other hand, the nonlinear activation functions, such as Sigmoid, ReLU, and tanh, have Lipschitz constants between 0 and 1. Computing an exact Lipschitz constant of an NN is, in general, challenging due to the piecewise application of the activation functions, which makes the computation of the overall gradient bound complex. In fact, the work presented in \cite{Soumya} presented a rather conservative estimate, taken to be the product of the largest absolute row sum norm of weights of each layer \cite{NNLipschitz}. In this work, we employ a semidefinite programming (SDP) based approach \cite{LipSDP} for estimating the Lipschitz constant for NNs that is much less conservative and remains computationally efficient.

A third enhancement we introduce is the way to impose a desired Lipschitz bound for an NN controller: We keep the last layer linear, free from activation function, and simply scale down the weights of the last layer proportionately so as to achieve a desired Lipschiz bound, without affecting the weights of the inner layers of an otherwise optimal NN. The results for a real-life example of a quadcopter control are presented in Section~\ref{sec4}.

\subsection{Sector bound of nonlinearity and parameter variation}
The system of (\ref{eq6}) under state-feedback control $u(s)=\pi(s)=\pi_0(s)+\pi_{NN}(s)=Ks+u_{NN}(s)$ can be decomposed into linear and nonlinear and parameter variation (NPV) components, by linearization of the nominal system around the equilibrium and considering the rest of the nonlinearity:
\begin{align} 
\dot{s} &= f(s,u,\alpha)
\!+\!\!\left[\underbrace{J_{f,s}}_{A_0}s\!+\!\underbrace{J_{f,u}}_{B_0}\underbrace{u}_{Ks}\right]_{\colvec{s=0\\u_{NN}=0\\\alpha=0}}\!\!\!\!\!\!\!\!-\!(A_0s+B_0Ks) \nonumber \\ 
&=\underbrace{(A_0+B_0K)}_{A_K}s + \underbrace{f(s,u(s),\alpha)-(A_0+B_0K)s}_{{\rm NPV:} \zeta(s,u_{NN}(s),\alpha)}, \nonumber
\end{align}
\noindent where $J_{f,s}$ and $J_{f,u}$ are the Jacobian matrices of $f$ with respect to $s$ and $u$, respectively, and the overall NPV part is the component not included in the linearized nominal part, namely:  
\begin{align} 
&\zeta(s,u_{NN}(s),\alpha) = f(s,Ks+u_{NN}(s),\alpha)-A_Ks. \nonumber
\end{align}

The sector bounds of the NPV component over the state and input domains $\mathcal S,\mathcal U_{L,\mathcal S}$, respectively, are then derived as the bounds for the Jacobians of the NPV, $\zeta$:
\begin{align}
&\underline{\mathcal{L}}^{i,j} \leq J_{\zeta,s}^{i,j} \leq \bar{\mathcal{L}}^{i,j}, \forall i,j \in \{1,...,n\}  \nonumber \\
&\underline{\mathcal{L}}^{i,j+n} \leq J_{\zeta,u}^{i,j} \leq \bar{\mathcal{L}}^{i,j+n}, \forall i \in \{1,...,n\},  \forall j \in \{1,...,m\}. \nonumber 
\end{align}
These sector bounds then imply the existence of another QC as formulated in \cite[Proposition 2]{Soumya}.

\subsection{Lyapunov-based LMI for robust stability}
Using the QCs for the NN controller Lipschitz bound and system NPV sector bound, in \cite{Soumya}, an LMI constraint is obtained to guarantee the existence of a robust Lyapunov function over a given state domain, Lipschitz constant, and parameter variation bound. The feasibility of this LMI constraint ensures the existence of a robust quadratic Lyapunov function to certify the the robust stability of the NN-controlled system and a RoS within the user-specified safety set that is invariant and ensures convergence to equilibrium, even under parameter variations. \\   

\noindent \textbf{Theorem 1 (Linear Matrix Inequality Constraint \cite{Soumya})} \\  
\noindent For the given system of (\ref{eq6}), when the below LMI constraint of (\ref{eq9}) is feasible for some $L,K,(P \in \mathbb{S}^{+}),(\Lambda\geq 0),(\gamma_{ij}\geq 0,\forall i\in 1,\ldots,m,j\in 1,\ldots,n)$, then there exists a robust quadratic Lyapunov function $\mathcal{V}(s) = s^TPs$ such that $\pi(s)=Ks+\pi_{NN}(s)$ with $\pi_{NN}\in\Pi_L$ stabilizes the system of (\ref{eq6}) over the given state domain under the variations of the parameters in its given domain:
\begin{align}  
\left[\begin{array}{ccc} V_{L,\{\Gamma_j\},P}&*&*\\ \boldsymbol{0}_{m,n\times n}&M_{\chi\Lambda}-diag(\{\gamma_{ij}\})&*\\N_{s\Lambda}^T+R^T.P&N_{\chi\Lambda}^T&M_{\xi\Lambda} \end{array}\right] \prec 0, \label{eq9}
\end{align}
\\
\noindent where $\Gamma_{j} := \Sigma_{i=1}^m\gamma_{ij}$ for all $j \in \{1,...,n\}$,  and the matrices are defined as follows:  
\begin{align}  
&V_{L,\{\Gamma_j\},P}=M_{s\Lambda}+L^2.diag(\{\Gamma_j\})+PA_K+A_K^TP; \nonumber \\ 
&M_{s\Lambda} := diag\!\! \left(\sum_{i=1}^n \Lambda^{k_{ij}}\left( \bar{c}_{ij}^2 - c_{ij}^2 \right) | j \in 1,...,n\! \right);  \nonumber \\   
&k_{ij} := i + (j-1)n;  \nonumber \\
&c_{ij} := (\underline{\mathcal{L}}^{i,j} + \bar{\mathcal{L}}^{i,j}) /2;\quad \bar{c}_{ij} := max(|\underline{\mathcal{L}}^{i,j}|, |\bar{\mathcal{L}}^{i,j}|); \nonumber \\
&M_{\chi\Lambda}\! := Q^{T}\!diag\!\!\left(\sum_{i=1}^n \Lambda^{k_{ij}}\left( \bar{c}_{ij}^2 - c_{ij}^2 \right)\! | j \in n+1,...,n+m\! \right)\!\!Q; \nonumber \\
&Q := \boldsymbol{I}_m \odot \boldsymbol{1}_{1 \times n};\quad R := \boldsymbol{I}_n \odot \boldsymbol{1}_{1 \times (n+m)}; \nonumber \\ 
&M_{\xi\Lambda} := diag(-\Lambda); \nonumber \\ 
&N_{s\Lambda} := \left[ D_{s,1},...,D_{s,n} \right]; \nonumber \\ 
&D_{s,i} := \left[\begin{array}{ll} diag(\Lambda^{k_{ij}}.c_{ij} | j \in 1,...,n) & \boldsymbol{0}_{n \times m} \end{array} \right]; \nonumber \\
&N_{\chi\Lambda} := Q^T. \left[ D_{\chi,1}, ..., D_{\chi,n} \right]; \nonumber \\
&D_{\Lambda,i} := \left[ \boldsymbol{0}_{m \times n} diag(\Lambda^{k_{ij}}.c_{ij} | j \in n+1,...,n+m) \right]. \nonumber 
\end{align}

\section{Robust, Optimal, Safe, Stabilizing Control for Quadcopter}\label{sec4}
\subsection{Quadcopter dynamics \& its Nominal control}
We consider the Iris quadcopter \cite{IrisDrone} for control design, where Table \ref{Table 1} lists its parameter values.   

\begin{figure} [h!]
\vspace*{-.2in}
\centerline{\includegraphics[width=0.5\textwidth]{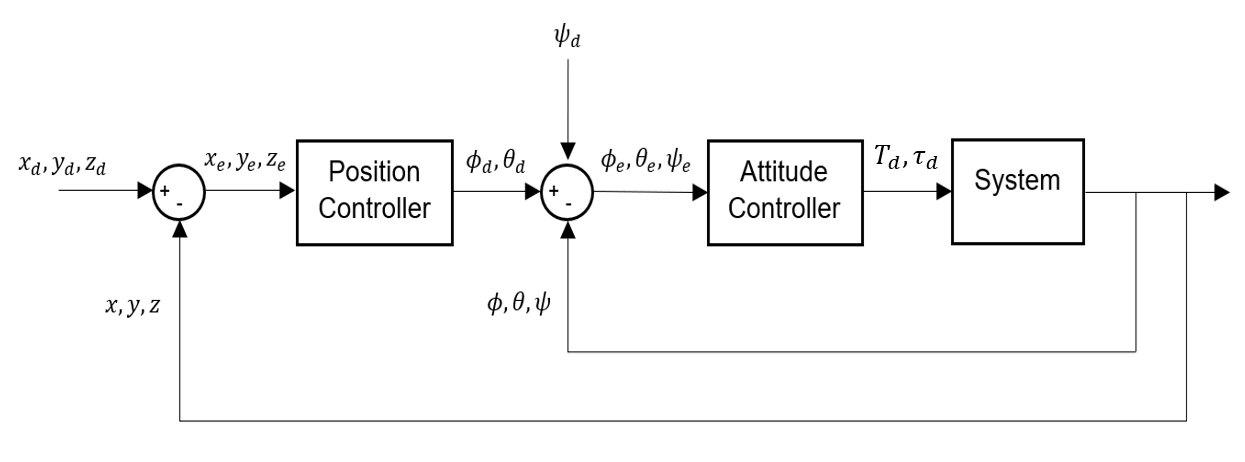}}
\vspace*{-.1in}
\caption{Nominal controller architecture}
\label{Nominal_controller}
\end{figure}

\begin{table}[h!]
\caption{Quadcopter model parameters}
\label{Table 1}
\vspace*{-.1in}
    \centering
    \resizebox{\linewidth}{!}{
\renewcommand{\arraystretch}{2.0}
\begin{tabular}{M{0.15\linewidth}|M{0.5\linewidth}|M{0.35\linewidth}}
\hline
\hline    
$I_{xx}, I_{yy}$ & x/y-axis rotational moment of inertia   & $2.9125\times10^{-2} \ [kg \cdot m^2]$  \\ \hline
$I_{zz}$ & z-axis rotational moment of inertia             & $5.5225\times10^{-2} \ [kg \cdot m^2]$  \\ \hline
$g$      & gravitational acceleration                      & $9.807 \ [m/s^2]$                       \\ \hline
$m$      & mass                                            & $1.5 \ [kg]$                            \\ \hline
$l$      & distance between center and rotor               & $0.25554 \ [m]$                         \\ \hline
$k_d/k_t$      & ratio between drag and lift constants            & $0.06$                                  \\ \hline
$D_{x},D_{y},D_{z}$  & x/y/z-axis drag coefficient         & $0.25$                                  \\ 
\hline
\hline
\end{tabular}}
\end{table}
A quadcopter is an underactuated system where the degree of freedom for control is less than the number of states, and a typical control specification comprises a desired position ($x_d,y_d,z_d$) and a desired heading ($\psi_d$). We employ a cascade proportional-derivative (PD) controller as the nominal controller in our design. Using the position error, the positional P-controller computes the desired thrust ($T_d$) and desired roll/pitch angles ($\phi_{d},\theta_{d}$), which are then used to determine the attitude error which the attitude PD-controller uses to calculate the desired torque ($\tau_d$). The structure of this nominal controller is shown in Fig \ref{Nominal_controller}.

The control inputs are set to be the motor thrusts:
\begin{align} 
u_i = k_t \omega_i^2, \forall i \in \{1,2,3,4\}, \nonumber
\end{align}
and these control inputs are allocated from the required thrust and torques using:
\begin{align}  
\left[\!\!\begin{array}{c} u_1\\u_2\\u_3\\u_4\end{array}\!\!\right] \!\!=\!\! 
\left[\begin{array}{cccc} 1&1&1&1\\-\frac{l}{\sqrt{2}}&-\frac{l}{\sqrt{2}}&\frac{l}{\sqrt{2}}&\frac{l}{\sqrt{2}}\\-\frac{l}{\sqrt{2}}&\frac{l}{\sqrt{2}}&\frac{l}{\sqrt{2}}&-\frac{l}{\sqrt{2}}\\k_d/k_t&-k_d/k_t&k_d/k_t&-k_d/k_t\end{array}\right]^{-1}\!\!\left[\!\!\begin{array}{c} T_d\\\tau_{\phi,d}\\\tau_{\theta,d}\\\tau_{\psi,d}\end{array}\!\!\right].  \nonumber
\end{align}

\noindent \textbf{For Position Control:}
\begin{align} 
\begin{array} {l} 
x_e = x_d - x, y_e = y_d - y, z_e = z_d - z; \\ 
\phi_d = -K_{y}y_{e}, \theta_d = K_{x}x_{e}, 
T_d = K_z  z_e.
\end{array} \nonumber
\end{align}

\noindent \textbf{For Attitude Control:}  
\begin{align}
\begin{array} {ll} 
\phi_e & = \phi_d - \phi = -K_{y}y_{e} - \phi; \\
\theta_e & = \theta_d - \theta = K_{x}x_{e} - \theta; \\
\psi_e&=\psi_{d}-\psi;\\
\tau_{\phi,d} & = K_{\phi}\phi_{e}+K_{\dot{\phi}}\dot{\phi}_{e};  \\
              & = K_{\phi}[-K_{y}(y_{d}-y)-\phi] + K_{\dot{\phi}}[K_{y}\dot{y} -\dot{\phi}]; \\
\tau_{\theta,d} & = K_{\theta}\theta_{e}+K_{\dot{\theta}}\dot{\theta}_{e}; \\ 
                & = K_{\theta}[K_{x}(x_{d}-x)-\theta] + K_{\dot{\theta}}[-K_{x}\dot{x} - \dot{\theta}]; \\
\tau_{\psi,d} & =  K_{\psi}\psi_{e}+K_{\dot{\psi}}\dot{\psi}_{e} = K_{\psi}(\psi_{d}-\psi) - K_{\dot{\psi}}\dot{\psi}.
\end{array} \nonumber
\end{align}

The thrust and torque generated by the quadcopter's rotors are proportional to the superposition of the square of rotor speeds. However, these values are influenced by aerodynamic conditions and external disturbances \cite{AerodynamicEffect}. To account for these uncertainties, a parameter variation is introduced into the thrust and torque model that is then compensated through the added NN controller. 

Table \ref{Table 3} represents the corresponding parameter variation bounds, indicating $\pm 5\%$ variation in the actual thrust and torque from their commanded values. Modeling these variations ensures that the model accounts for the real-world discrepancies arising from aerodynamic conditions, enhancing the robustness of the control design.  

\begin{table}[h!]
\caption{Quadcopter parameter variation bounds}
\label{Table 3}
\vspace*{-.1in}
    \centering
    \resizebox{\linewidth}{!}{
\renewcommand{\arraystretch}{2.0}
\begin{tabular}{M{0.2\linewidth}|M{0.4\linewidth}|M{0.4\linewidth}}
\hline
\hline
$\alpha_{1}$ & Thrust variance          & $[-0.05, \ 0.05]$  \\ \hline
$\alpha_{2}$ & Roll torque variance     & $[-0.05, \ 0.05]$  \\ \hline
$\alpha_{3}$ & Pitch torque variance    & $[-0.05, \ 0.05]$  \\ \hline
$\alpha_{4}$ & Yaw torque variance      & $[-0.05, \ 0.05]$  \\ 
\hline
\hline
\end{tabular}}
\end{table}

Therefore, the quadcopter dynamics with parameter variation and nominal controller becomes: 
\begin{align} f(s,u,\alpha) & =\left[\!\! \begin{array}{cc} \dot{x} \\ 
                                                            \dot{y} \\   
                                                            \dot{z} \\ 
                                                            p+S_{\phi}T_{\theta}q+C_{\phi}T_{\theta}r \\ 
                                                            C_{\phi}q-S_{\theta}r \\ 
                                                            \frac{S_{\phi}}{C_{\theta}}q+\frac{C_{\phi}}{C_{\theta}}r \\ 
                                                            \frac{(1+\alpha_{1})T}{m}(C_{\psi}S_{\theta}C_{\phi}+S_{\psi}S_{\phi}) - \frac{D_{x}}{m}\dot{x} \\
                                                            \frac{(1+\alpha_{1})T}{m}(S_{\psi}S_{\theta}C_{\phi}-C_{\psi}S_{\phi}) - \frac{D_{y}}{m}\dot{y} \\ 
                                                            -g+\frac{(1+\alpha_{1})T}{m}(C_{\theta}C_{\phi}) - \frac{D_{z}}{m}\dot{z} \\ 
                                                            \big((I_{yy}-I_{zz})qr+(1+\alpha_{2})\tau_{\phi}\big)/I_{xx} \\
                                                            \big((I_{zz}-I_{xx})pr+(1+\alpha_{3})\tau_{\theta}\big)/I_{yy} \\
                                                            \big((I_{xx}-I_{yy})pq+(1+\alpha_{4})\tau_{\psi}\big)/I_{zz} \\ 
                                                             \end{array}\!\! \right], \label{eq10}
\end{align}
where $T = mg - K_{z}z +u_{1}+u_{2}+u_{3}+u_{4}$; \\
      $\tau_{\phi} = K_{\phi}(K_{y}y-\phi) + K_{\dot{\phi}}[K_{y}\dot{y} -\dot{\phi}] + \frac{l}{\sqrt{2}}(-u_{1}-u_{2}+u_{3}+u_{4}) $;\\
      $\tau_{\theta} = -K_{\theta}(K_{x}x+\theta) - K_{\dot{\theta}}(K_{x}\dot{x} + \dot{\theta}) +\frac{l}{\sqrt{2}}(-u_{1}+u_{2}+u_{3}-u_{4}) $;\\
      $\tau_{\psi} = -K_{\psi}\psi - K_{\dot{\psi}}\dot{\psi} + b(u_{1}-u_{2}+u_{3}-u_{4}) $. \\ 
\\
The system equation (\ref{eq10}) consists of 12 states, 4 control inputs, and 4 parameter variations. The coordinates are shifted such that the equilibrium point is at the origin, i.e., $(x_d,y_d,z_d,\psi_d) = (0,0,0,0)$. With this coordinate shift, so that $(x_d,y_d,z_d,\psi_d) = (0,0,0,0)$, it follows that nominal Position and Attitude controllers are state-feedback. 

Table \ref{Table 2} lists the parameter values for the nominal controller, which is chosen to ensure that the state matrix associated with the nominal controller, $A_K$, is stable with all its eigenvalues having negative real parts, i.e., the nominal controller stabilizes the linearized nominal system under zero parameter variation around the equilibrium point. 
\begin{table}[h!]
\caption{Nominal PD controller gains}
\label{Table 2}
\vspace*{-.1in}
    \centering
    \resizebox{\linewidth}{!}{
\renewcommand{\arraystretch}{2.0}
\begin{tabular}{M{0.2\linewidth}|M{0.6\linewidth}|M{0.2\linewidth}}
\hline
\hline
$K_{x}, K_{y}$         & P gain of x/y-direction      & 0.05    \\ \hline
$K_{z}$                & P gain of z-direction        & 0.1    \\ \hline
$K_{\phi}, K_{\theta}$ & P gain of $\phi$/$\theta$ rotation & 0.1    \\ \hline
$K_{\psi}$           & P gain of $\psi$ rotation    & 0.1    \\ \hline
$K_{\dot{\phi}}, K_{\dot{\theta}}$ & D gain of $\phi$/$\theta$ rotation    & 0.01    \\ \hline
$K_{\dot{\psi}}$     & D gain of $\psi$ rotation    & 0.1    \\ 
\hline
\hline
\end{tabular}}
\end{table}

\subsection{Maximal Lipschitz bound \& Invariant safety domain}
The feasibility of the LMI constraint (\ref{eq9}) depends on system nonlinearity, safety domain, and parametric bound, and when feasible, one can find $(L,K,P,\Lambda,\{\lambda_{ij}\})$ that satisfy the LMI constraint so that the system (\ref{eq6}) under the control $\pi(s)=Ks+\pi_{NN}(s);\,\pi_{NN}\in\Pi_L$ is certified as asymptotically stable with the Lyapunov function $\mathcal{V}=s^TPs$ having a RS-RoS (Robust Safe Region of Stability) that is a sublevel set of $\mathcal{V}$. To maximize the class of robust, safe, and stable NN controllers, $L$ should be as large as possible, and to maximize the size of RS-RoS, the corresponding safety domain $\mathcal{S}$ should also be as large as possible. Maximal ($L,\mathcal{S}$) values are determined through the iterative search method of \cite[Algorithm 1]{Soumya}. The process begins with an initial guess ($L_0,\mathcal{S}_0$) set to small values, which for our quadcopter case is set as below: 
$$ L_0 = 1.0; \mathcal{S}_0\!\equiv\! \left[\!\! \begin{array}{c} \mathcal{S}_0(x)=\mathcal{S}_0(y)=\mathcal{S}_0(z) \\\mathcal{S}_0(\phi)=\mathcal{S}_0(\theta)=\mathcal{S}_0(\psi) \\ \mathcal{S}_0(\dot{x})=\mathcal{S}_0(\dot{y})=\mathcal{S}_0(\dot{z}) \\ \mathcal{S}_0(p)=\mathcal{S}_0(q)=\mathcal{S}_0(r) \end{array} \!\!\right] \!=\! \left[\!\! \begin{array}{c} (-1.5,1.5) \\ (-0.2,0.2) \\ (-1.5,1.5) \\ (-0.2,0.2) \end{array} \!\!\right]\!. $$

The ($L,\mathcal{S}$) values are iteratively increased, and the feasibility of the LMI constraint (\ref{eq9}) is evaluated at each step. If the LMI constraint becomes infeasible, ($L,\mathcal{S}$) is reduced, and the feasibility is rechecked. This process is repeated with progressively smaller step sizes until the algorithm converges to the maximal Lipschitz bound and safety domain, denoted as ($L^*,\mathcal{S}^*$), which in our case is found to be:
$$ L^* \!=\! 1.2613; \mathcal{S}^* \!\!\equiv\!\!\! \left[\!\!\!\! \begin{array}{c} \mathcal{S}^*(x)\!=\!\mathcal{S}^*(y)=\!\mathcal{S}^*(z) \\ \mathcal{S}^*(\phi)\!=\!\mathcal{S}^*(\theta)\!=\!\mathcal{S}^*(\psi) \\ \mathcal{S}^*(\dot{x})\!=\!\mathcal{S}^*(\dot{y})=\!\mathcal{S}^*(\dot{z}) \\ \mathcal{S}^*(p)\!=\!\mathcal{S}^*(q)\!=\!\mathcal{S}^*(r) \end{array}\!\!\!\! \right] \!\!\!=\!\!\! \left[\!\!\!\! \begin{array}{c} (-1.892,1.892) \\ (-0.252,0.252) \\ (-1.892,1.892) \\ (-0.266,0.266) \end{array} \!\!\!\!\right]\!\!. $$

\subsection{Optimal NN controller training}
To train an optimal NN controller satisfying the maximal Lipschitz bound $L^*$ found above so as to minimize reference tracking error and control cost, we employ Proximity Policy Optimization (PPO) \cite{PPO}, a deep reinforcement learning (RL) method based on an actor-critic structure. In an RL, an agent interacts with the environment to collect samples, where each sample is represented as $(s,u,r,\gamma, s')$ where $s \in \mathcal{S}, u \in \mathcal{U} $ are the current state and action ($\mathcal{S}$ and $\mathcal{U}$ are state and action domains), $r \in \mathbb{R}$ is the associated reward, $\gamma \in [0,1)$ is a discount factor, and  $s'\in \mathcal{S}$ is the next state. These samples represent the transition from $s$ to $s'$ when the action $u$ is taken and the reward $r$ is received. Using the collected samples, policy and value, the neural networks used as function approximators, representing maps from state to action (actor network) and from state to value (value network) are trained, where value is the expected accumulated total rewards. (In contrast, the reward is an immediate return of a single action step.) The goal of RL is to optimize the network parameters to maximize the value.

\begin{figure}[htbp]
\vspace*{-.2in}
\centerline{\includegraphics[width=0.5\textwidth]{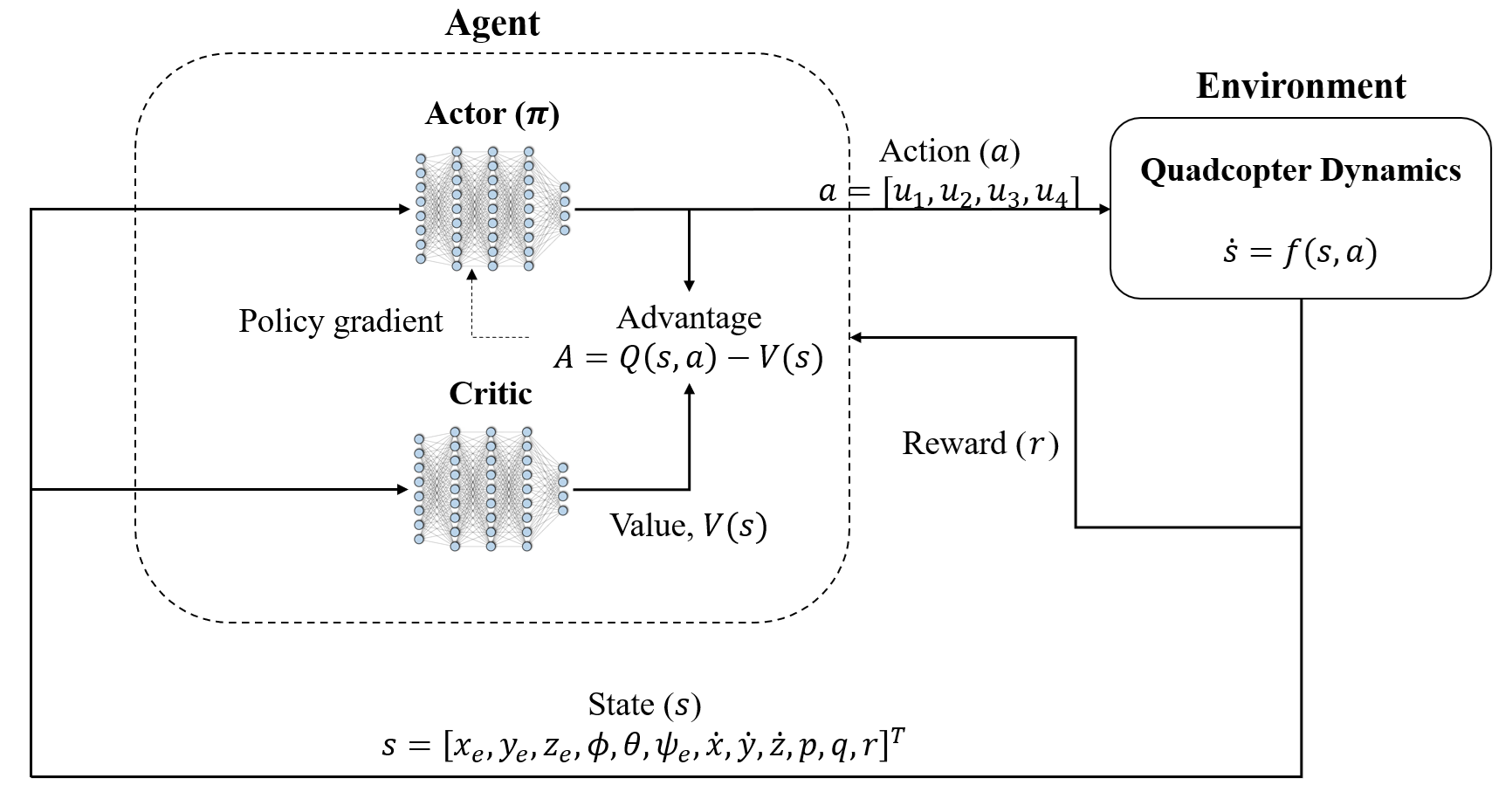}}
\caption{Actor Critic Framework for RL training}
\label{Fig.3}
\end{figure}

Fig \ref{Fig.3} illustrates this actor-critic framework used in our training of optimal NN controller. The environment models the quadcopter dynamics along with the nominal controller described in Equation (\ref{eq10}). The NN controller provides control inputs that are fed into the environment (nominally controlled quadcopter), which executes the control and returns a ``reward" value depending on the current state, and also advances the state. The ``advantage" is defined as the difference between the state value and an action value, providing a measure of whether a particular action is better or worse than the default action suggested by the policy. We estimated the advantage using the method proposed in \cite{GAE}, and computed the policy gradient based on this estimated advantage.

The reward indicates how effectively the agent performs in achieving the desired objective, guiding the optimization of the control policy. 
In our case, the reward function minimizes the error between the reference state and the current state. When devising the control action, both the transient and steady-state responses are factored in: When the quadcopter is far from the reference position, its reward is computed from only the positional and directional errors. Once the quadcopter is close to the reference position, angular velocity, which must become small, is also factored in along with the positional and directional errors so as to reduce the oscillations around the reference. Given the position and direction errors defined as $p_{e} := [x_d-x,y_d-y,z_d-z]^T$ and $\psi_{e}:=\psi_d-\psi$, and the angular velocity defined by $\nu := [p,q,r]^T$, the reward function is formulated as:
\begin{align} r = \bigg \lbrace\begin{array}{ll} 0.8r_{pos} + 0.2r_{\psi}                 & p_e > 0.3 \\                                
                                                    0.6r_{pos} + 0.2r_{\psi} + 0.2r_{\nu} & {\rm otherwise}
                                                    \end{array} \nonumber  
\end{align}

\noindent where $r_{pos} = e^{-2\norm{p_e}}, r_{\psi} = e^{-|\psi_e|}, r_{\nu} = e^{-2\norm{\nu}}$.

\begin{figure}[htbp]
\vspace*{-.1in}
\centerline{\includegraphics[width=0.5\textwidth]{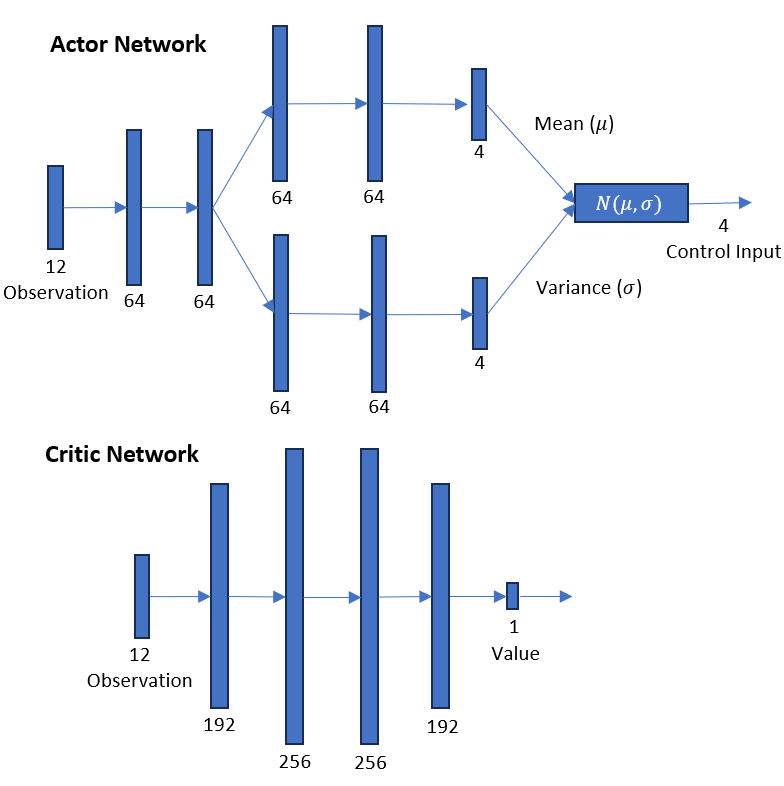}}
\caption{Actor and Critic networks for quadcopter control}
\label{Fig.4}
\end{figure}
Fig \ref{Fig.4} shows the actor and critic network structures. In our design, the actor consists of five hidden layers, each containing 64 neurons with Rectified Linear Unit (ReLU) as the activation function. The final layer performs only linear operations without an activation function. The actor has two output branches: one computes the mean, and the other computes the variance of the action. Actions are then sampled from a multivariate normal distribution using the mean and variance during the training phase. During the testing phase, only the mean values are used to control the quadcopter. The critic network consists of four hidden layers, each with 192 or 256 neurons and ReLU activation function. This network evaluates the value function, guiding the optimization of the actor network.

At the beginning of each episode, the quadcopter's position is initialized at the origin, and the reference position is randomly selected within a two-meter radius from the origin. Parameter variation ($\alpha$) is randomly sampled at every timestep from an uniform distribution within the given bounds, introducing uncertainties in thrust and torque. The simulation uses the first-order integration to update the state based on (\ref{eq10}). 
Both actor and critic networks are trained using PPO. Table~\ref{Table 4} provides the PPO parameters we used for NN training. Definitions of these parameters can be found in \cite{PPO}, \cite{GAE}, and \cite{TRPO}. 

\begin{table}[htbp]
\caption{PPO parameters used in Reinforcement Learning}
\label{Table 4}
    \centering
    \resizebox{\linewidth}{!}{
\renewcommand{\arraystretch}{2.0}
\begin{tabular}{M{0.3\linewidth}|M{0.2\linewidth}||M{0.3\linewidth}|M{0.2\linewidth}}
\hline
\hline
Clip Factor           & 0.2   & Actor learning rate  & 0.001 \\ \hline
Discount Factor       & 0.99  & Critic learning rate & 0.01  \\ \hline
GAE Factor            & 0.95  & Experience horizon   & 2048  \\ \hline
Entropy loss weight   & 0.4   & Minibatch size       & 2048   \\
\hline
\hline
\end{tabular}}
\end{table}

\subsection{Certifying robust safety and stability}
Robust safety is certified simply by starting within the maximal RS-ROS set $\mathcal{S}^*$ contained in the user-specified safety set, which by virtue of being a sublevel set of the computed robust Lyapunov function is invariant under system evolution, including random parametric variation, thereby preserving safety. Next, to certify the robust asymptotic stability of the closed-loop system, the Lipschitz constant of the NN controller must remain below the maximal computed Lipschitz bound $L^*$. It is therefore important to obtain a tight estimation of an NN's Lipschitz constant. To achieve this, other than introducing L2-norm for Lipschitz continuity, we apply semi-definite programming based method \cite{LipSDP} to estimate the Lipschitz constant of the trained actor network. In case the estimated Lipschitz bound of the trained NN is higher than the $L^*$ value, we simply scale down the weights of the final layer of the actor network (which by design is purely linear with no activation) so that the maximal Lipschitz bound requirement is satisfied. In our study, the Lipschitz constant of the trained actor network came out to be $L=3.1519>L^*=1.2613$. So we scaled down the weights of the final layer by a factor of $0.4<L^*/L=0.4002$. After the corresponding scaling down, the Lipschitz bound of the NN controller is reduced to: 
\begin{align}
L^{new} = 1.2608<L^*=1.2613. \nonumber
\end{align}
Note since in our design, the final layer is a linear operation without an activation function, scaling it down does not alter the action pattern inferred by the actor network. Thereby this adjustment retains the NN controller's functional behavior while also certifying its robust stability. The reduced conservativeness from the use of i) L2-norm for Lipschitz continuity, together with ii) semi-definite programming to estimate the Lipschitz bound of a NN, is demonstrated through the $L$ values in Table \ref{Table 5}, showing $>\!\!2400$-fold improvement.

\begin{table}[htbp]
\caption{Estimates of Lipschitz bound $L$}
\label{Table 5}
    \centering
    \resizebox{\linewidth}{!}{
\renewcommand{\arraystretch}{2.0}
\begin{tabular}{M{0.5\linewidth}|M{0.5\linewidth}}
\hline
\hline
$L$ from $\infty$-norm \cite{NNLipschitz} & $L$ from L2-norm and using SDP\cite{LipSDP}  \\ \hline
2920 & 1.2608                          \\ 
\hline
\hline
\end{tabular}}
\end{table}

\section{Evaluation of Synthesized Controller}
We evaluate the performance of the controlled quadcopter under the nominal and NN controllers designed above. During the simulation of the controlled quadcopter, the $\alpha$ parameters are randomly drawn from the uniform distribution within the specified parameter bounds. 
\begin{figure} [ht!]
\vspace*{-.1in}
    \centering
    \begin{subfigure}{0.48\linewidth}
        \centering
        \includegraphics[width=4.8cm]{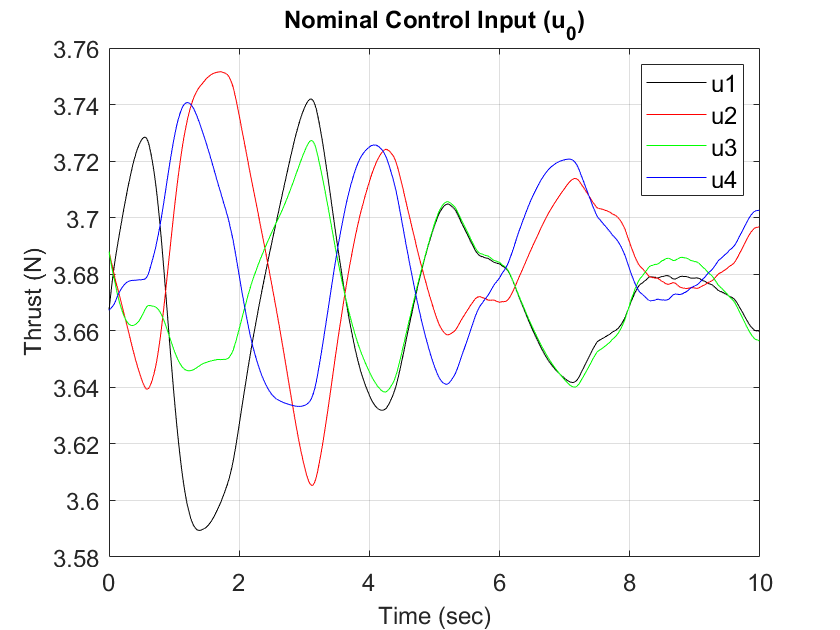}
        \caption{Nominal Control Input $u_0$}
        \label{fig: NominalControlInput}
    \end{subfigure}
    \begin{subfigure}{0.48\linewidth}
        \centering   
        \includegraphics[width=4.8cm]{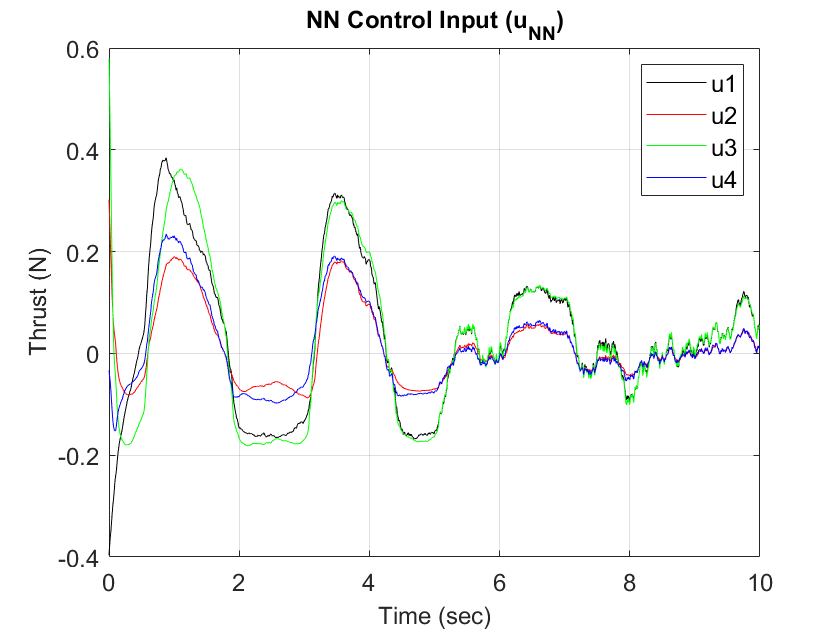}
        \caption{NN Control Input $u_{NN}$}
        \label{fig: NNControlInput}
    \end{subfigure}
    \hfill
    \begin{subfigure}{0.48\linewidth}
        \centering   
        \includegraphics[width=4.8cm]{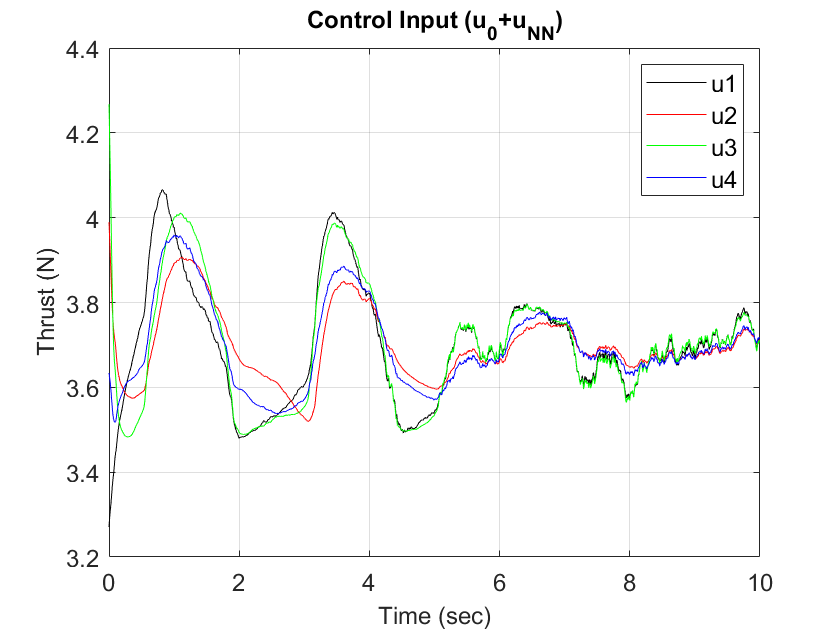}
        \caption{Control Input $u = u_0+u_{NN}$}
        \label{fig: ControlInput}
    \end{subfigure}
\caption{Reference position control input}\label{Fig.5}
\end{figure}

\begin{figure*}[ht!]
    \centering
    \begin{subfigure}{0.49\linewidth}
        \centering
        \includegraphics[width=9.0cm, height=7cm]{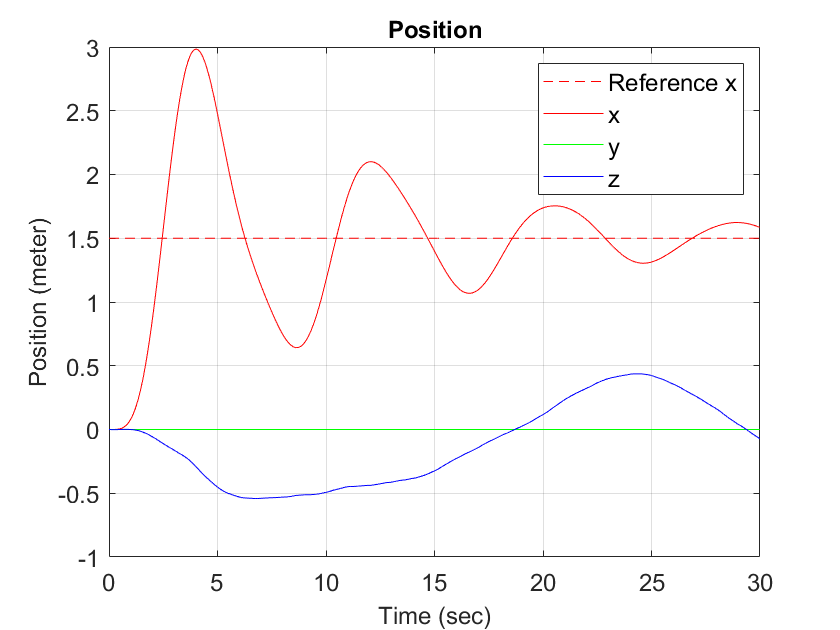}
        \caption{Position with $\pi = \pi_0$}
        \label{Fig.6(a)}
    \end{subfigure}
    \begin{subfigure}{0.49\linewidth}
        \centering   
        \includegraphics[width=9.0cm, height=7cm]{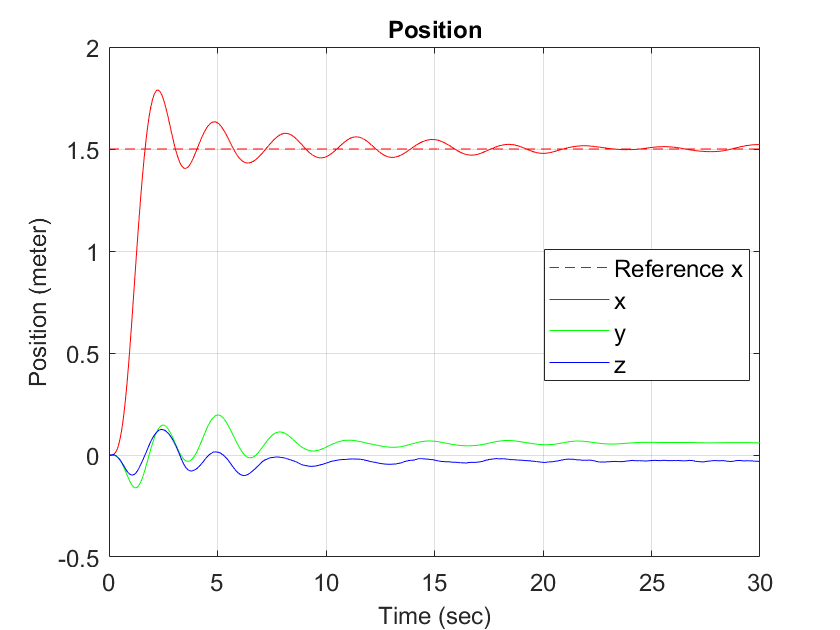}
        \caption{Position with $\pi = \pi_0+\pi_{NN}$}
        \label{Fig.6(b)}
    \end{subfigure}
    \hfill
    \begin{subfigure}{0.49\linewidth}
        \centering
        \includegraphics[width=9.0cm, height=7cm]{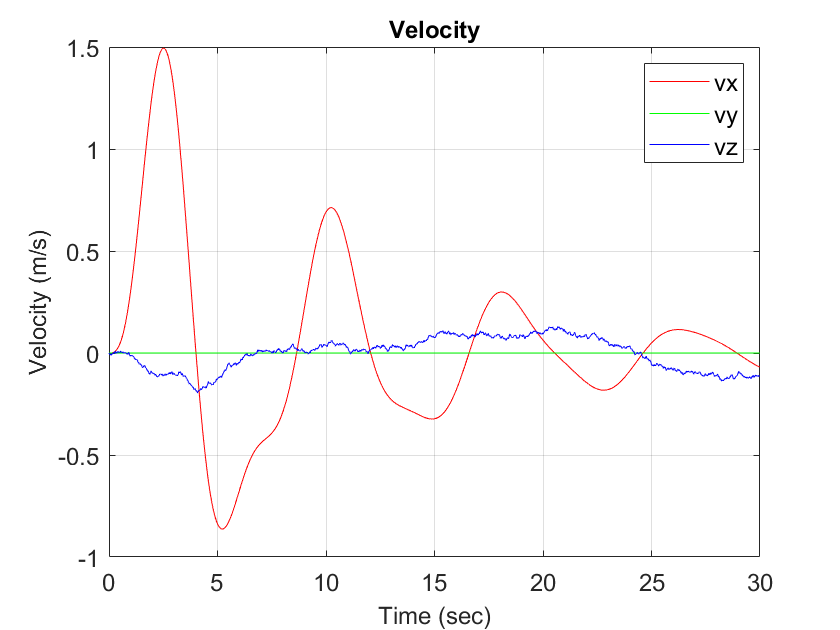}
        \caption{Velocity with $\pi = \pi_0$}
        \label{Fig.6(c)}
    \end{subfigure}
    \begin{subfigure}{0.49\linewidth}
        \centering   
        \includegraphics[width=9.0cm, height=7cm]{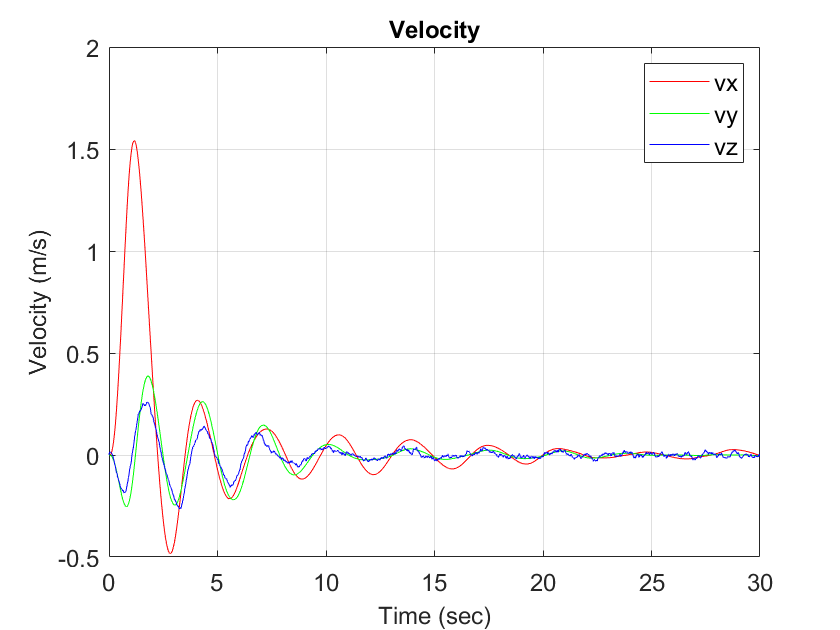}
        \caption{Velocity with $\pi = \pi_0+\pi_{NN}$}
        \label{Fig.6(d)}
    \end{subfigure}
    \hfill
    \begin{subfigure}{0.49\linewidth}
        \centering
        \includegraphics[width=9.0cm, height=7cm]{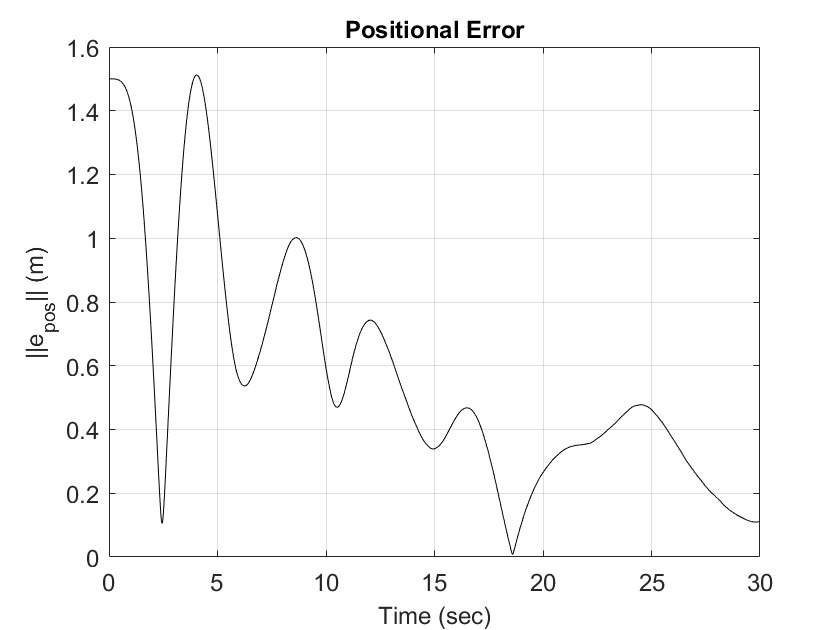}
        \caption{Positional error with $\pi = \pi_0$}
        \label{Fig.6(e)}
    \end{subfigure}
    \begin{subfigure}{0.49\linewidth}
        \centering   
        \includegraphics[width=9.0cm, height=7cm]{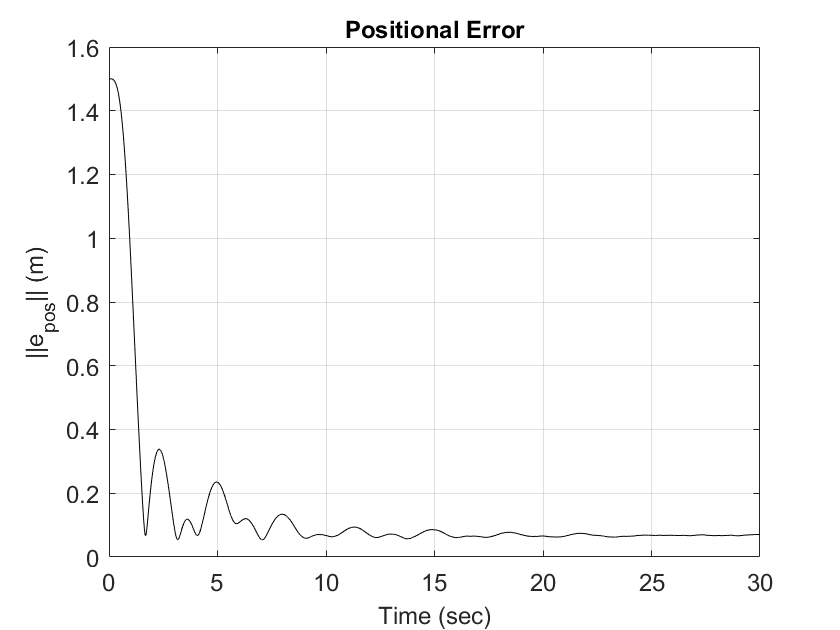}
        \caption{Positional error with $\pi = \pi_0+\pi_{NN}$}
        \label{Fig.6(f)}
    \end{subfigure}
\caption{Results of reference position tracking}
\label{Fig.6}
\end{figure*}

\begin{figure*}[ht!]
    \centering
    \begin{subfigure}{0.49\linewidth}
        \centering
        \includegraphics[width=9.0cm]{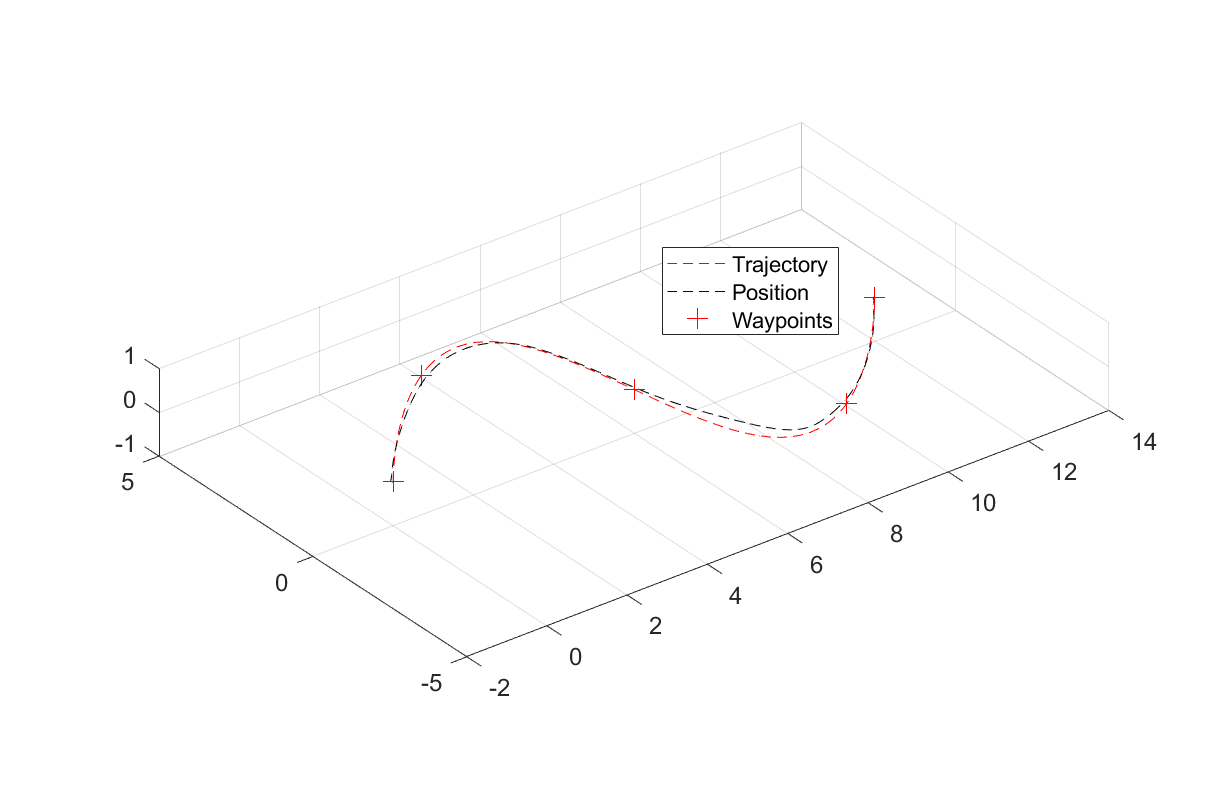}
        \caption{Trajectory}
        \label{Fig.7(a)}
    \end{subfigure}
    \begin{subfigure}{0.49\linewidth}
        \centering   
        \includegraphics[width=9.0cm]{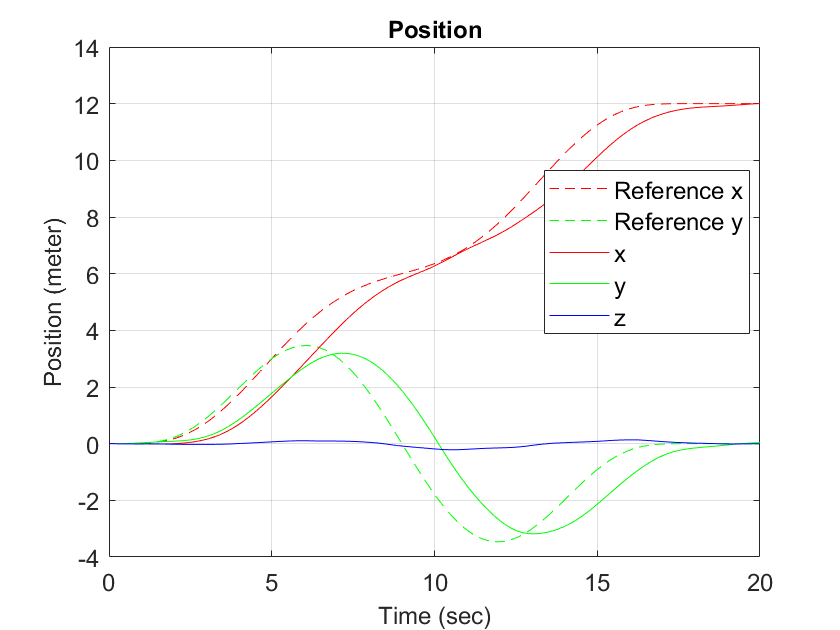}
        \caption{Position}
        \label{Fig.7(b)}
    \end{subfigure}
    \hfill
    \begin{subfigure}{0.49\linewidth}
        \centering
        \includegraphics[width=9.0cm]{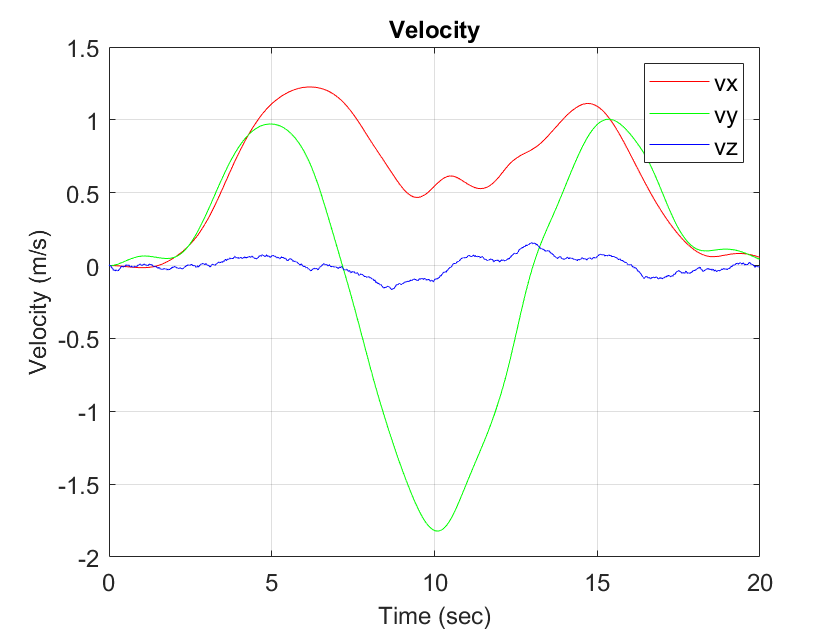}
        \caption{Velocity}
        \label{Fig.7(c)}
    \end{subfigure}
    \begin{subfigure}{0.49\linewidth}
        \centering   
        \includegraphics[width=9.0cm]{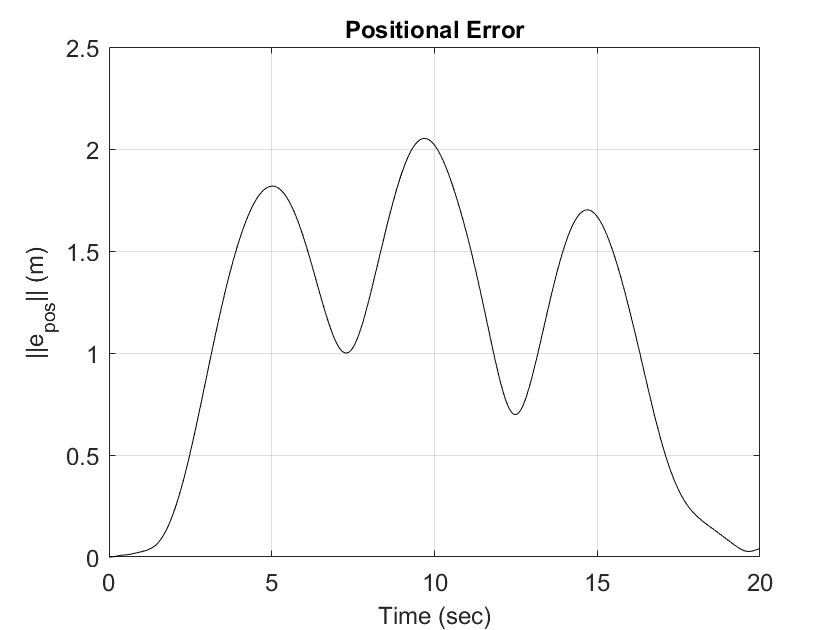}
        \caption{Positional Error}
        \label{Fig.7(d)}
    \end{subfigure}
\caption{Results of reference trajectory tracking}
\label{Fig.7}
\end{figure*}

\subsection{Reference position tracking}
The reference position tracking capability of the quadcopter was evaluated under two scenarios: (1) using only the nominal controller and (2) using both the nominal and NN controllers. The quadcopter was initialized at $p_{init}=[0,0,0]^T$ with zero initial attitude, linear and angular velocities, while the reference linear position was set to $p_{d}=[1.4,0,0]^T$. 

Control inputs from the nominal controller, the NN controller, and their combined values are shown in Figure \ref{Fig.5}. These inputs highlight the NN controller's role in dynamically adjusting the system's behavior to minimize position errors.

Figure \ref{Fig.6} demonstrates the controller performance in terms of quadcopter position, velocity, and positional error. In the nominal controller-only scenario (Figure \ref{Fig.6}(a), (c), (e)), the quadcopter successfully stabilized at the reference point but with slower error reduction. When combined with the NN controller (Figure \ref{Fig.6}(b), (d), (f)), the positional error decreased much more rapidly, indicating improved transient performance. The NN controller effectively augmented the nominal controller by providing additional control inputs, leading to enhanced tracking accuracy and reduced response time, even under parameter variations. 

\subsection{Reference trajectory tracking}
To evaluate reference trajectory tracking performance, user-specified timed waypoints on an $S$-shaped trajectory were used as the initial specification, as listed in Table \ref{Table 5}. A path-planning algorithm of the minimum snap method \cite{Snap} was then used to map the timed waypoints of Table \ref{Table 5} to generate a smooth reference trajectory as a discrete time function for position, velocity, acceleration, and direction---See the red trajectory in Figure~\ref{Fig.7}(a). 
During the control, the target position at each discrete sample time was set to be the path planner's computed position for the next sample instant, and our designed controller was employed for computing the revised control action by simply changing the target position from that of the current sample instant to that of the next sample instant.

\begin{table}[h!]
\caption{Timed waypoints for a reference trajectory}
\label{Table 6}
    \centering
    \resizebox{\linewidth}{!}{
\renewcommand{\arraystretch}{2.0}
\begin{tabular}{M{0.2\linewidth}|M{0.6\linewidth}|M{0.2\linewidth}}
\hline
\hline
Waypoint      & Position      & Time    \\ 
\hline
\hline
$p_1$         & $[0, 0, 0]^T$     & 0  \\ \hline
$p_2$         & $[3, 3, 0]^T$     & 5  \\ \hline
$p_3$         & $[6, 0, 0]^T$     & 9  \\ \hline
$p_4$         & $[9, -3, 0]^T$    & 13 \\ \hline
$p_5$         & $[12, 0, 0]^T$    & 18 \\
\hline
\hline
\end{tabular}}
\end{table}

Figure~\ref{Fig.7} illustrates the quadcopter's performance in tracking the $S$-shaped trajectory, where the reference vs. the actual trajectories are shown in red vs. black in Figure~\ref{Fig.7}(a). The quadcopter successfully followed the reference trajectory by maintaining positional accuracy: Positional error remained low throughout the trajectory, validating the effectiveness of the NN controller in maintaining trajectory tracking precision even under parameter variations. The result demonstrates that the proposed control design is robust to environmental uncertainties and capable of trajectory tracking. 

\section{Conclusion}
This paper proposed novel improvements to the Robust Optimal Safe and Stability Guaranteed Control (ROSS-GC) method of \cite{Soumya}, which enabled it for a successful first-time application for the control of a quadcopter for reference trajectory tracking in the presence of thrust and torque uncertainties, as is the case with the real-world scenarios. The developed controller with the improved design methodology showed the feasibility of a robust, optimal, safe, and stability-guaranteed controller synthesis for a quadcopter for the first time. The optimality stems from the minimization of trajectory deviation. The method involved combining a designed nominal cascaded PD controller with a NN controller---The former was designed to stabilize the nominal linearized system under no parametric changes, whereas the added NN control stabilized the given nonlinear system with parametric uncertainty. The addition of the NN controller demonstratively enhanced the transient/steady-state responses, showing its capability to handle parameter variations effectively, while the use of the robust stability-guaranteed approach certified the system's asymptotic stability in spite of the parameter variations within the specified bounds.
Future work can examine extending the method to more complex UAV models and validation through high-fidelity simulations as well as real-world testing, thereby further validating the general applicability of the proposed robust, optimal, safe, and stability-guaranteed NN controllers in practical environments.

\bibliographystyle{Bibliography/IEEEtranTIE}
\bibliography{reference}

\end{document}